\documentclass[aps,pra,twocolumn,amsmath,amssymb,nofootinbib,showpacs,superscriptaddress]{revtex4-1}
\usepackage[english]{babel}
\usepackage{latexsym}
\usepackage{graphics}
\usepackage{graphicx}
\usepackage{epsfig}
\usepackage{color}
\usepackage{bm}
\usepackage{amsmath}
\usepackage{amssymb}
\usepackage{amsthm}
\usepackage{dcolumn}
\usepackage{bm}
\usepackage{float}
\usepackage{hyperref}
\usepackage{color}
\usepackage{epstopdf}
\usepackage{cleveref}
\usepackage[svgnames]{xcolor}
\hypersetup{hidelinks,colorlinks=true,allcolors=DarkBlue}

\usepackage[ruled, vlined]{algorithm2e}
\usepackage[normalem]{ulem}

\newcommand{\uv}{\ensuremath{{\bf u}}}

\newcommand{\vv}{\ensuremath{{\bf v}}}

\newcommand{\wv}{\ensuremath{{\bf w}}}

\newcommand{\xv}{\ensuremath{{\bf x}}}
\newcommand{\xestv}{\ensuremath{{\hat{\bf x}}}}

\newcommand{\erestv}{\ensuremath{{\hat{\bf e}}}}
\newcommand{\yv}{\ensuremath{{\bf y}}}
\newcommand{\zv}{\ensuremath{{\bf z}}}

\newcommand{\Gm}{\ensuremath{{\bf G}}}
\newcommand{\Hm}{\ensuremath{{\bf H}}}

\newcommand{\Kb}{\ensuremath{\mathcal{K}}}

\newcommand{\xextv}{\ensuremath{{\bf x}^{\rm ext}}}
\newcommand{\yextv}{\ensuremath{{\bf y}^{\rm ext}}}
\newcommand{\Rbase}{\ensuremath{R_{\rm base}}}

\begin{document}

\preprint{APS/123-QED}

\title{Blind information reconciliation with polar codes for quantum key distribution}
	
	\author{E.O. Kiktenko}
	\affiliation{Russian Quantum Center, Skolkovo, Moscow 143025, Russia}
	\affiliation{Moscow Institute of Physics and Technology, Dolgoprudny 141700, Russia}
	\affiliation{	Department of Mathematical Methods for Quantum Technologies, Steklov Mathematical Institute of Russian Academy of Sciences, Moscow 119991, Russia}
	\affiliation{NTI Center for Quantum Communications, National University of Science and Technology MISiS, Moscow 119049, Russia}

	\author{A.O. Malyshev}
	\affiliation{Russian Quantum Center, Skolkovo, Moscow 143025, Russia}
	\affiliation{Department of Physics, University of Oxford, Oxford OX1 3PG, UK}

	\author{A.K. Fedorov}
	\affiliation{Russian Quantum Center, Skolkovo, Moscow 143025, Russia}
	\affiliation{Moscow Institute of Physics and Technology, Dolgoprudny 141700, Russia}

	\date{\today}
	\begin{abstract}
	We suggest a new protocol for the information reconciliation stage of quantum key distribution based on polar codes.
	The suggested approach is based on the blind technique, which is proved to be useful for low-density parity-check (LDPC) codes.
	We show that the suggested protocol outperforms the blind reconciliation with LDPC codes, especially when there are high fluctuations in quantum bit error rate (QBER).
	\end{abstract}

\maketitle

\section{Introduction}
Quantum key distribution (QKD) allows growing a secret key between two legitimate users connected by a quantum and authenticated classical channels~\cite{Gisin2002,Scarani2009,Lo2014,Lo2016}.
The security of QKD is based on the laws of quantum physics and it is guaranteed to be secure against any unforeseen technological developments, such as quantum computing~\cite{Shor1997}. 

A workflow of QKD devices can be divided into two phases~\cite{Gisin2002,Scarani2009}. 
During the first phase, QKD devices encode information in quantum bits (qubits), transmit and measure them, and then discard the records about preparation and measurement events occurred in incompatible bases.
As a result of this phase, two legitimate parties, Alice and Bob, obtain so-called sifted keys, which are highly correlated but not identical bit strings.
The measure of discrepancy between sifted keys is characterized by the quantum bit error rate (QBER).
By virtue of quantum mechanics, QBER allows parties to estimate amount of information intercepted by an adversary Eve in the first phase.
If it is low enough, parties proceed to the second phase, where they strive to distill a pair of secure keys out from the sifted keys.
This phase consists of (i) information reconciliation procedure, where Alice and Bob communicate via the public authenticated channel to remove discrepancies in the sifted keys; 
(ii) privacy amplification, where parties shorten their keys to eliminate the information which Eve has obtained by then.

Information reconciliation methods proposed so far mainly rely on the Cascade method~\cite{Brassard1994,Martinez2015,Pedersen2015} 
or low-density parity-check (LDPC) codes~\cite{Elkouss2009,Martinez2010,Martinez20102,Martinez2012,Mink2012,Martinez2013,Kiktenko2017,Liu2020}.
The latter approach has been studied in various aspects and it is used in commercial QKD devices.  
An important LDPC-based information reconciliation technique is the blind method~\cite{Martinez2012}, which allows operating without an a priori estimation of QBER.
Its further improvement~\cite{Kiktenko2017} introduces symmetry into the reconciliation and thus provides high efficiency with small number of communication rounds between parties. 
We note that this method requires the same level of computational power for both sides of communications.
In addition, this protocol is applicable only for discrete variables (DV) QKD protocols: in the case of continuous variables (CV) QKD, a one-way reverse reconciliation is required~\cite{Scarani2009}.

However, the performance of LDPC-code-based reconciliation schemes substantially degrades in the case of highly fluctuating QBER.
As we demonstrate below, the probability of successful decoding for LDPC codes is a non-smooth function of disclosed bits number, which undermines the efficiency of the reconciliation.
To overcome this challenge we construct a novel blind scheme based on polar codes.
To the best of our knowledge, these codes were previously considered for non-interactive protocols only~\cite{Nakassis2014,Jouguet2014,Nakassis2017,Yan2018,Lee2018,Lee2018b,Arikan2008}.
We also show that even though in non-interactive protocols polar codes demonstrate performance similar to that of LDPC codes, in the interactive regime they show a strong advantage, especially in the case of highly fluctuating QBER.
To validate relevance of such consideration, we provide experimental evidence for high fluctuations of the QBER in real channels.

\section{LDPC-code-based blind reconciliation} \label{sec:LDPC}

Information reconciliation based on LDPC codes works as follows.
Before start of QKD phases, Alice and Bob agree upon some LDPC code with $M\times N$ parity-check matrix \Hm{}. 
At the beginning of the reconciliation stage, Alice calculates a syndrome ${\bf s}=\Hm\xv$ for her sifted key \xv{} of length $N$ and discloses it over the public channel 
(all operations are performed modulo-2, and both syndrome and keyzsare treated as column-vectors).
Then Bob employs his $N$-bit sifted key \yv{} and announced syndrome ${\bf s}$ as input to the decoder.
If decoding fails, parties drop this pair of keys and proceed to the next one.
If decoding succeeds, parties verify reconciled keys by comparing their $\varepsilon$-almost universal$_2$ 
(for details, see Ref.~\cite{Fedorov2018}).
If the verification check succeeds, parties put \xv{} and \xestv{} to their storages of the verified key.
Otherwise, they behave as if decoding has failed.

The main figure of merit for successful reconciliation is called efficiency and is expressed as follows:
\begin{equation}\label{equ:ldpc_straight_eff}
	f = \frac{M}{N \cdot h_{\rm bin}(q)} = \frac{1 - R}{h_{\rm bin}(q)}.
\end{equation}
Here $R=1-M/N$ is the code rate of the employed LDPC code, $q$ is the QBER, and $h_{\rm bin}(q)=-q\log_2q-(1-q)\log_2(1-q)$ is the binary entropy function.
According to the Slepian-Wolf bound~\cite{Slepian1973}, decoding may succeed only when syndrome length is greater or equal than $N h_{\rm bin}(q)$.
Meanwhile, exactly $M$ bits leak to Eve, so it is preferable to keep syndrome length as small as possible.
Hence, the value of $f$ shows how superfluous was the information revealed via the public channel; the closer $f$ is to unity, the more efficient is the protocol.

{In practice reconciliation schemes are designed to cover a range of possible QBERs.
In this case use of only one code with fixed rate $R$ does not provide flexible enough reconciliation.
For lower $q$ it results in excessive amount of disclosed information, while for higher QBERs it decreases the probability of successful decoding.
To provide smooth rate adaption one may use a pool of codes, as well as shortening and puncturing techniques widely used in coding theory~\cite{Martinez2010}.}
Shortened bits are ones at pre-agreed positions with fixed known values, and punctured are ones at pre-agreed positions with truly random values.
If $s$ and $p$ are numbers of shortened and punctured bits correspondingly, the resulting rate reads as $R = (N-M-s)/(N-s-p)$.
We note that usually the total number of auxiliary bits $d = s+p$ is fixed.
In further text to define $d$ we use a parameter $\alpha$ such that $d = N\alpha$.

However, the problem of rate-adaption is still relevant to the case when the actual value of the QBER is unknown, e.g. due to fluctuations of the quantum channel quality.
To address this issue, it was proposed to adopt an analogue of the hybrid automatic repeat request scheme (HARQ) from the field of wireless communications~\cite{Martinez2012}.
The workflow of the resulting protocol, known as blind reconciliation, is as follows.

{\it Step 0.} Alice and Bob choose an LDPC code with $M\times N$ parity check matrix \Hm{} and rate $\Rbase=1-M/N$. 
Parties also choose $\alpha$, which defines the number of auxiliary bits, and $\delta$, which defines the number of bits disclosed in each additional round (we assume that $\delta$ divides $\alpha N$).

{\it Step 1.} Alice and Bob take sifted keys $\xv$ and $\yv$ of length $(1-\alpha)N$ and extend them with $\alpha N$ punctured bits.
Let us denote the resulting $N$-bit extended key of Alice (Bob) by $\xextv{}$ ($\yextv{}$).
The positions for puncturing are usually chosen with a pseudo-random generator (PRG) initialized with the same seed on both sides. 

{\it Step 3.} Alice sends to Bob syndrome ${\bf s}=\Hm \xextv{}$.

{\it Step 4.} Bob tries to perform syndrome decoding and reports to Alice whether he succeeded or failed.
In the case of success, parties complete the protocol and proceed to the verification stage, otherwise, parties proceed to Step 5.

{\it Step 5.} If the remaining number of punctured bits is greater or equal to $\delta$, Alice turns $\delta$ punctured bits into shortened by sending Bob values of these $\delta$ bits, and parties go back to the Step 4 (the positions for newly shortened bits are chosen with synchronized PRG). 
Otherwise, parties decide that the whole blind reconciliation protocol failed.

According to the Ref.~\cite{Martinez2012}, the efficiency is as follows:
\begin{equation}\label{eq:LDPC-eff}
	f = \frac{M - \alpha N + n\delta}{N(1- \alpha) \cdot h_{\rm bin}(q)},
\end{equation}
where $n$ is the resulting number of additional communication rounds performed in Step 5.
We note that during the blind reconciliation the effective code rate $R$ varies in the range
\begin{equation} \label{eq:R-range-LDPC}
	R_{\max}^{\rm LDPC}:=\frac{\Rbase}{1-\alpha}\geq{R}\geq \frac{\Rbase-\alpha}{1-\alpha} =:R_{\min}^{\rm LDPC}
\end{equation}
(first, the rate $R_{\max}^{\rm LDPC}$ is tried, then it is reduced down to $R_{\max}^{\rm LDPC}-\delta/N$, $R_{\max}^{\rm LDPC}-2\delta/N$, and so on down to $R_{\min}^{\rm LDPC}$).

Performance comparison shows that this approach allows parties to have better rate-adaptability and thus to improve the efficiency of the reconciliation~\cite{Martinez2012}.
Note that the parameter $\alpha$, which is responsible for the rate-adaptability, should be chosen with care.
On the one hand, it determines the available range of rates in Eq.~\eqref{eq:R-range-LDPC}, and as it is increased, more code rates become available.
On the other hand, higher values of $\alpha$ lead to smaller blocks of sifted keys being taken during each stage, and therefore to the longer processing time of the whole keys.
Hence, we can introduce another way to widen the range of available rates by modifying Step 5 as follows.

{\it Step 5*.} If the remaining number of punctured bits is greater or equal to $\delta$, Alice turns $\delta$ punctured bits into shortened by sending Bob values of these $\delta$ bits.
Otherwise, Alice turns $\delta$ sifted key positions into shortened by disclosing them to Bob.
Then parties go to Step 4 
(the resulting efficiency is still given by Eq.~\eqref{eq:LDPC-eff}, whichever bits were disclosed in additional rounds).

\begin{figure}
    \centering
    \includegraphics[width=0.85\linewidth]{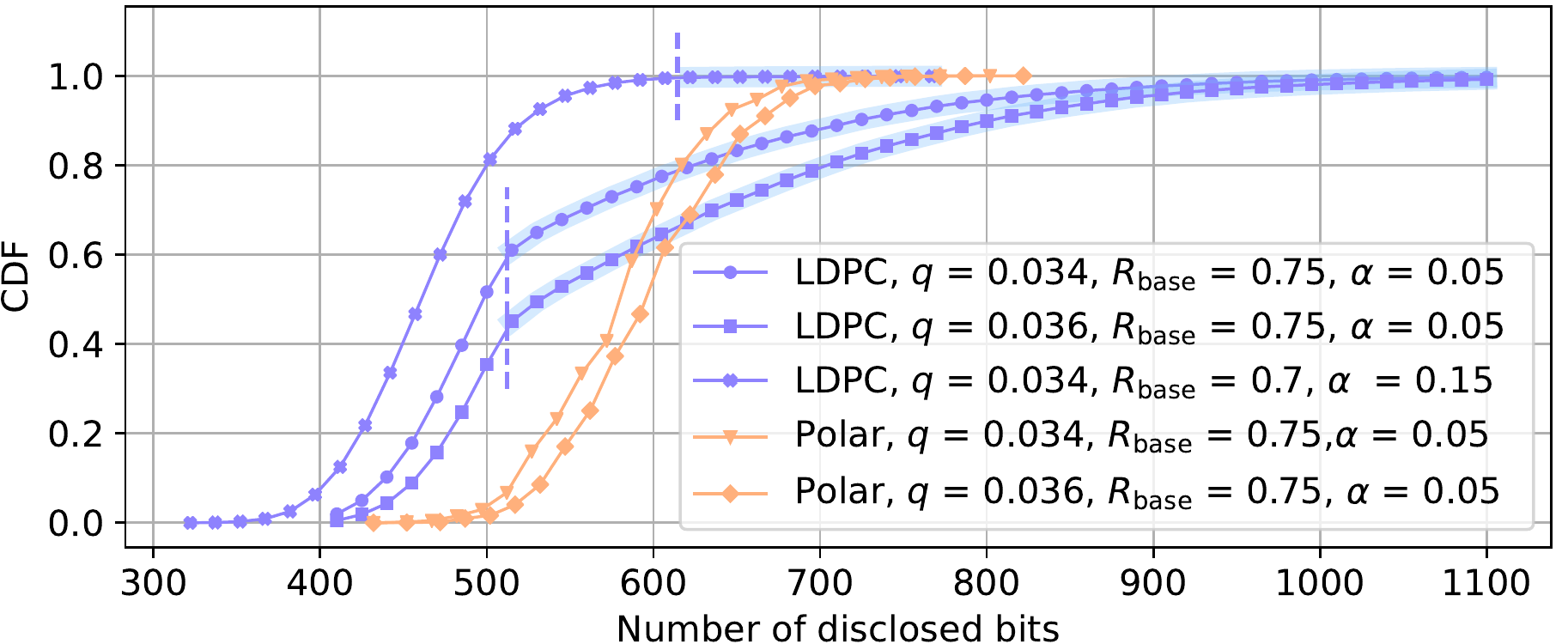}
    \vskip -2mm
    \caption{
    {CDFs for the number of disclosed bits required for successful reconciliation with blind protocols based on LDPC and polar codes.
    Vertical lines and highlighting indicate the point where all punctured bits are exhausted (in case of LDPC codes) and gain from each additional communication round drops down.
    The protocol based on polar codes does not suffer from this issue.}}
    \label{fig:CDF}
\end{figure}

The modified blind scheme is indeed able to cope with fluctuating QBER, but its performance deteriorates when fluctuations are too high.
Bits of Alice's sifted key have conditional entropy of $h_{\rm bin}(q) < 1$ with respect to the bits of Bob's sifted key, while truly random punctured bits have conditional entropy equal to $1$.
Hence, when sifted key bits are disclosed, the gain for successful decoding probability is smaller compared to the case of disclosed punctured bits.
This issue is captured by simulation results shown in Fig.~\ref{fig:CDF} (the results for polar codes are discussed in the next sections).
One can see that the cumulative distribution function (CDF) for the number of disclosed bits required for successful reconciliation has a breakpoint after all punctured bits are spent, and then it increases much slower with the disclosed bits number.
As a result, the efficiency degrades in the regions of QBER where parties have to disclose bits of sifted keys --- we observe this in our numeric experiments, which are discussed below.
The straightforward way to solve this issue is to increase the value of $\alpha$, and thus extend the range of achievable rates shown in Eq.~\eqref{eq:R-range-LDPC}.
However, this solution decreases the length of processed sifted keys $N(1-\alpha)$ and thus increases the total number of protocol runs necessary to reconcile a given block of sifted keys reducing net reconciliation throughput.
In what follows, we propose an original scheme based on polar codes which are free of this drawback.

\begin{figure*}
	\centering
	\includegraphics[height=0.24\linewidth]{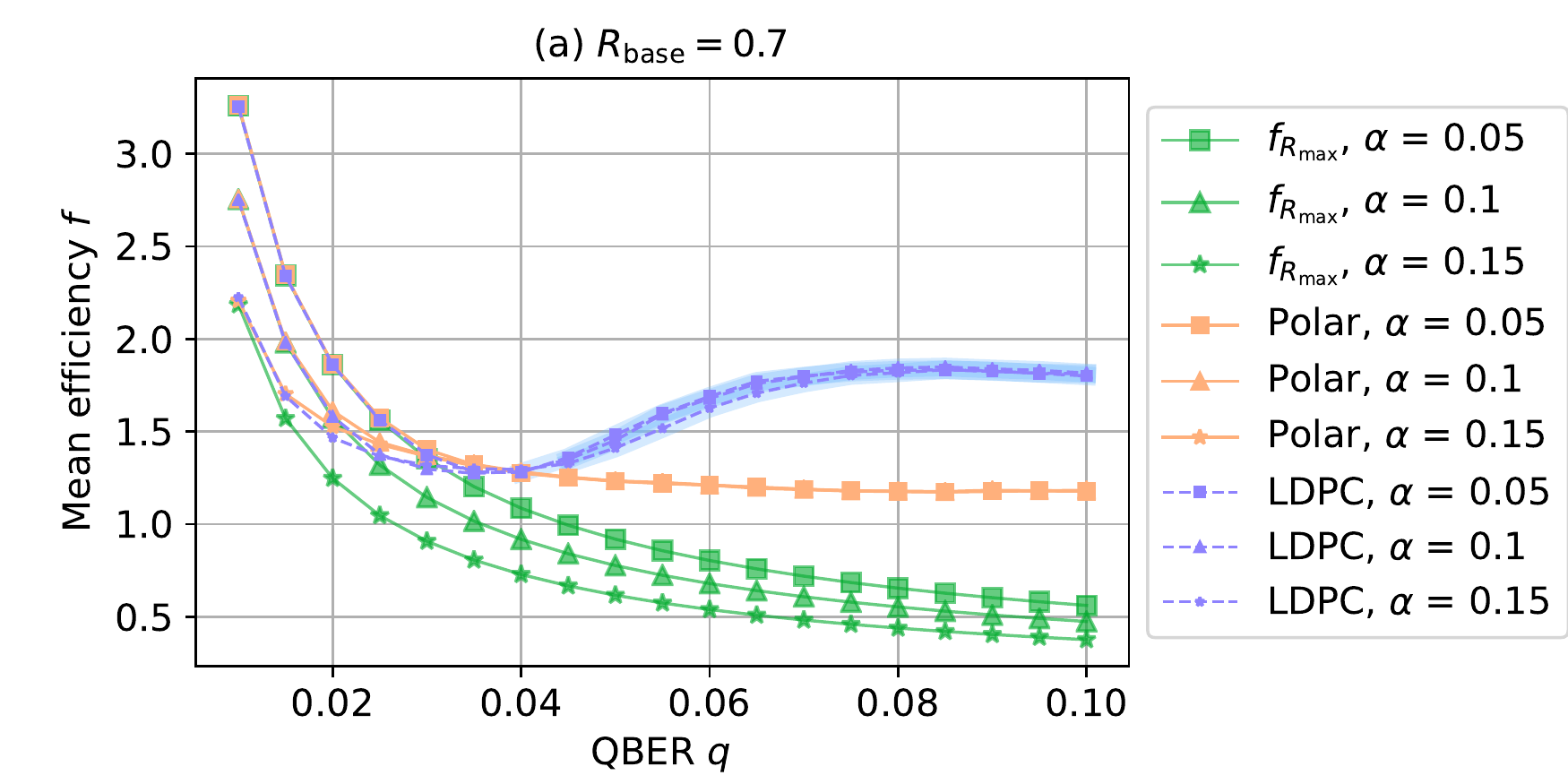}
	\includegraphics[height=0.24\linewidth]{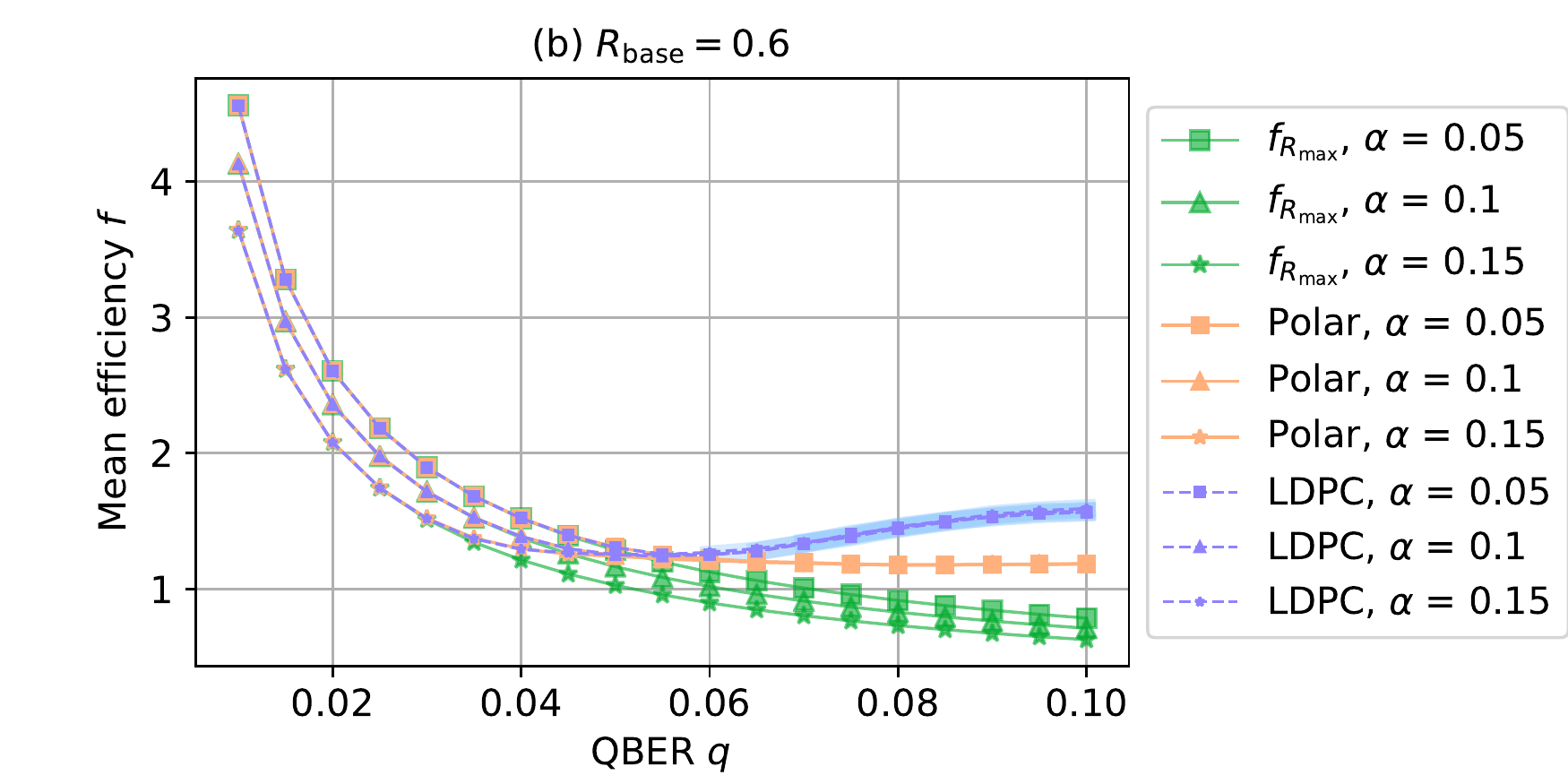} \\
	\includegraphics[height=0.24\linewidth]{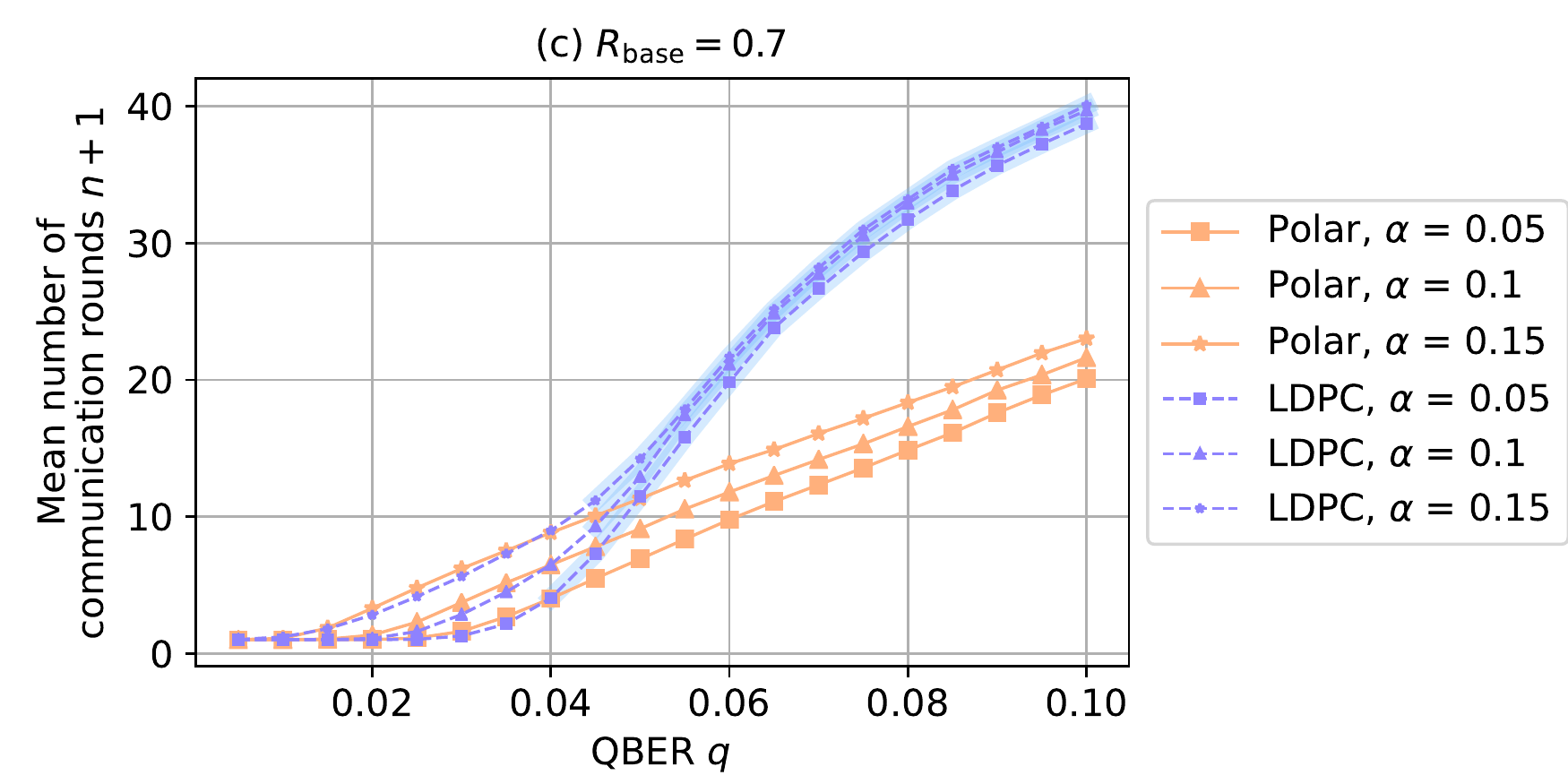}
	\includegraphics[height=0.24\linewidth]{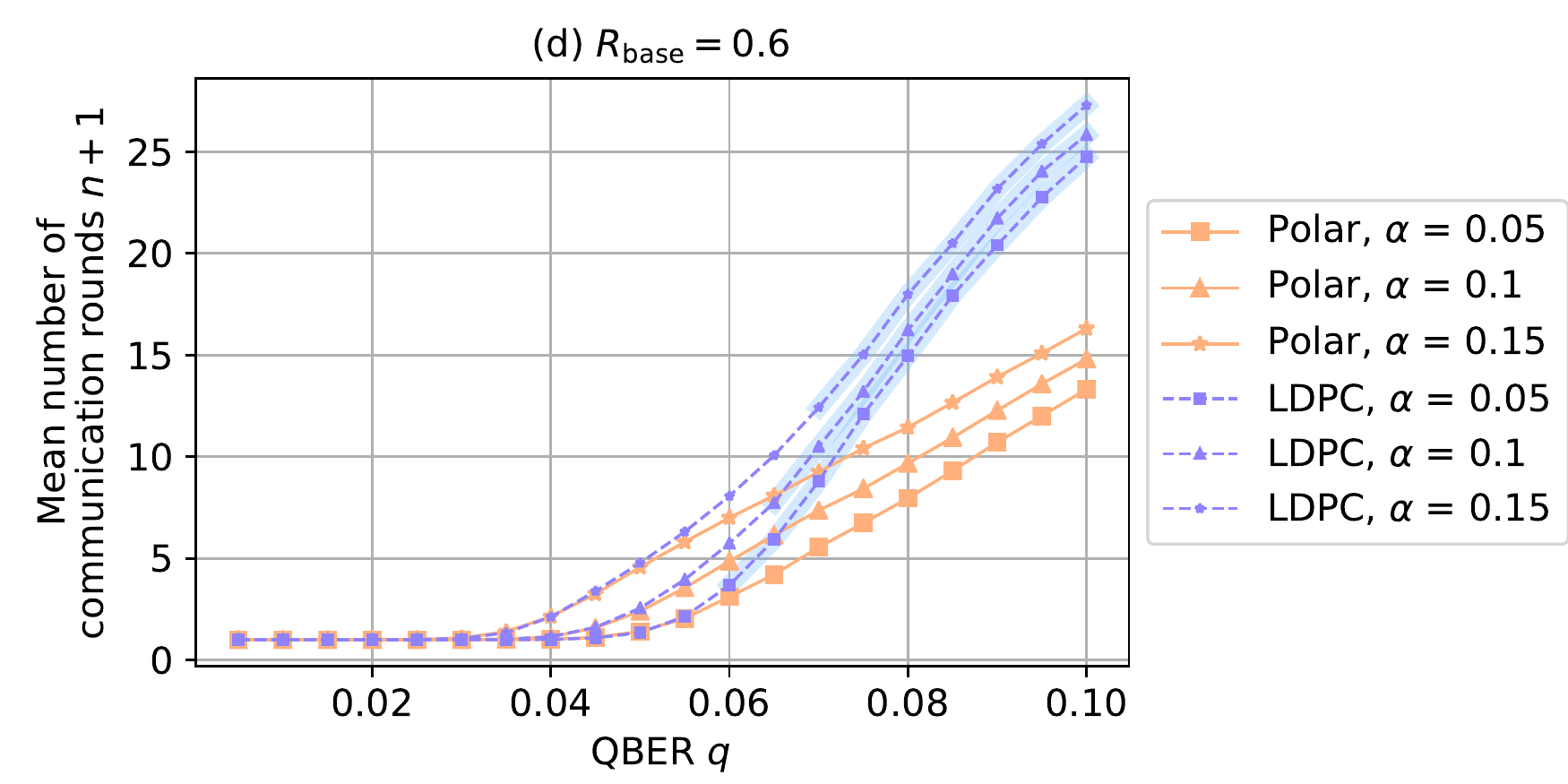}
	\vskip -2mm
	\caption{Comparison of mean efficiency (a, b) and mean number of communication rounds (c,d) for LDPC and polar codes-based blind reconciliation protocols with $\Rbase=0.7$ (a,c) and $\Rbase=0.6$ (b, d).
		For the LDPC codes-based protocols, the region where all punctured bits are spent is highlighted.}
	\label{fig:test-1}
\end{figure*}

\section{Polar code-based blind reconciliation}

Polar codes are linear error-correcting codes which operate as follows. 
Having $N=2^n$ (with $n>0$) copies of a real noisy channel $W$, one can consider $N$ specially constructed virtual channels, capacities of which tend to \textit{polarize}, i.e. to be close either to zero or to unity.
The fraction of virtual channels with close to unity capacity equals the capacity of real channel $C(W)$, meanwhile approximately $(1 - C(W)) \cdot N$ channels have capacity near zero.
Moreover, the higher is the value of $N$, the more accurate becomes this relation.

Information bits may be sent over $C(W) \cdot N$ channels, which are close to the ideal without any encoding.
The remaining bits are called \textit{frozen} and are commonly set to zero.
To encode information back from virtual channels to real, an $N$-bit vector consisting of information and frozen bits is multiplied by $N \times N$ generator matrix \Gm{}.
The basic problem of polar codes encoding is how to choose positions of frozen bits.
It turns out to be a hard problem since the direct calculation of each virtual channel capacity requires $\mathcal{O}(2^N)$ operations.
However, there are many heuristics developed, allowing to estimate the ``quality" of a virtual channel, and thus to choose the worst of them~\cite{Arikan2008,He2018}.
Whichever method is in use, its final aim is to derive an order of bit indices $\mathcal{K}$, arranged from the worst to the best.
Then first $C(W) \cdot N$ out of them are considered to be frozen ones, while others are information ones.

The use of CRC-aided Tal-Vardy list decoder significantly improves error correction ability of polar codes~\cite{Tal2015}.
In this approach a $c$-bit CRC of a transmitted $(K-c)$-bit message is written in $c$ out of $K$ non-frozen bits.
The list decoder keeps up to $L$ alternative decoding ``paths'' at each step of decoding and at the final step chooses output of that decoding path, which has correct CRC for decoded informational bits.
The decoding algorithm has complexity $\mathcal{O}(L \cdot N \log{N})$ and pushes error correction ability of polar codes to that of state-of-the-art LDPC and Turbo codes~\cite{Tal2015}.

Next we propose a polar codes-based blind reconciliation protocol, which is 
the main contribution of our work. 
It is inspired by the LDPC codes-based version and takes benefit of the fact that virtual channels are sorted with respect to their capacities, what we show with further numeric simulations.

{\it Step 0.} Alice and Bob choose a polar code with the rate $R_{\max}$ given by $R_{\max} = (K-c)/N$, where $N$, $K$, and $c$ are frame length, message length, and CRC-code length correspondingly.
Parties also agree on a number $\delta$ which defines the number of bits disclosed in each of additional communication rounds.

{\it Step 1.} Alice generates $K$ truly random information bits, calculates a $c$-bit CRC of them and composes a codeword \uv{}, with bits in first $(N - K - c)$ positions from \Kb{} set to zero, and informational plus CRC bits spread among remaining $K + c$ bits in a fashion known to Bob.

{\it Step 2.} Alice derives \vv{} --- a polar image of \uv{}, calculates $\wv = \xv \oplus \vv$, where  \xv{} is the Alice's sifted key of length $N$, and sends \wv{} to Bob via the public channel.

{\it Step 3.} Bob calculates $\zv = \wv \oplus \yv$, where \yv{} is Bob's sifted key corresponding to \xv{} and performs decoding, taking into account known  values of frozen bits.

{\it Step 4.} If upon the end of decoding, output of some decoding path contains a codeword with valid CRC, then the protocol is completed, Bob obtains the recycled key $\xestv{}=\yv \oplus \erestv$, where $\erestv$ is an obtained vector of errors, and parties proceed to the verification stage.
Otherwise, Bob reports failure to Alice and parties proceed to Step 5.

{\it Step 5} Alice announces additional $\delta$ bits from yet undisclosed bits. 
Their positions are $\delta$ ones to the right of the most right position of already disclosed bits in the ranking \Kb{}.
Then Step 4 is repeated.

According to the security analysis performed in Ref.~\cite{Lee2018} the overall efficiency reads as
\begin{equation}\label{eq:polar-eff}
	f = \frac{N - K + c + n\delta}{N \cdot h_{\rm bin}(q)},
\end{equation}
where $n$ is the resulting number of additional rounds.

An important difference in comparison with the LDPC codes-based version is that the proposed protocol has no limit on the number of additional communication rounds.
In the worst scenario all $N$ bits of \uv{} are disclosed.
During the protocol the effective code rate $R$ varies in range
\begin{equation} \label{eq:R-range-polar}
	R_{\max}^{\rm polar} := \frac{K-c}{N} \geq R \geq 0,
\end{equation}
so there is no any (unnatural) lower limit on the achievable rate.
Also we note that length of the processed sifted key in the protocol always equals to $N$ regardless of the covered code range.
These two features of the interactive polar codes-based protocol appear to be beneficial in the case of highly fluctuating QBER as we show in the next section.

\section{Performance analysis}\label{sec:analysis} 

To study the performance of the suggested protocol, we conduct two tests: with synthetic data and with experimental data from real QKD setup operating in urban environment.
In both tests, we compare the performance of LDPC and polar codes-based protocol.

\begin{figure*}
	\centering
	\includegraphics[height=0.25\linewidth]{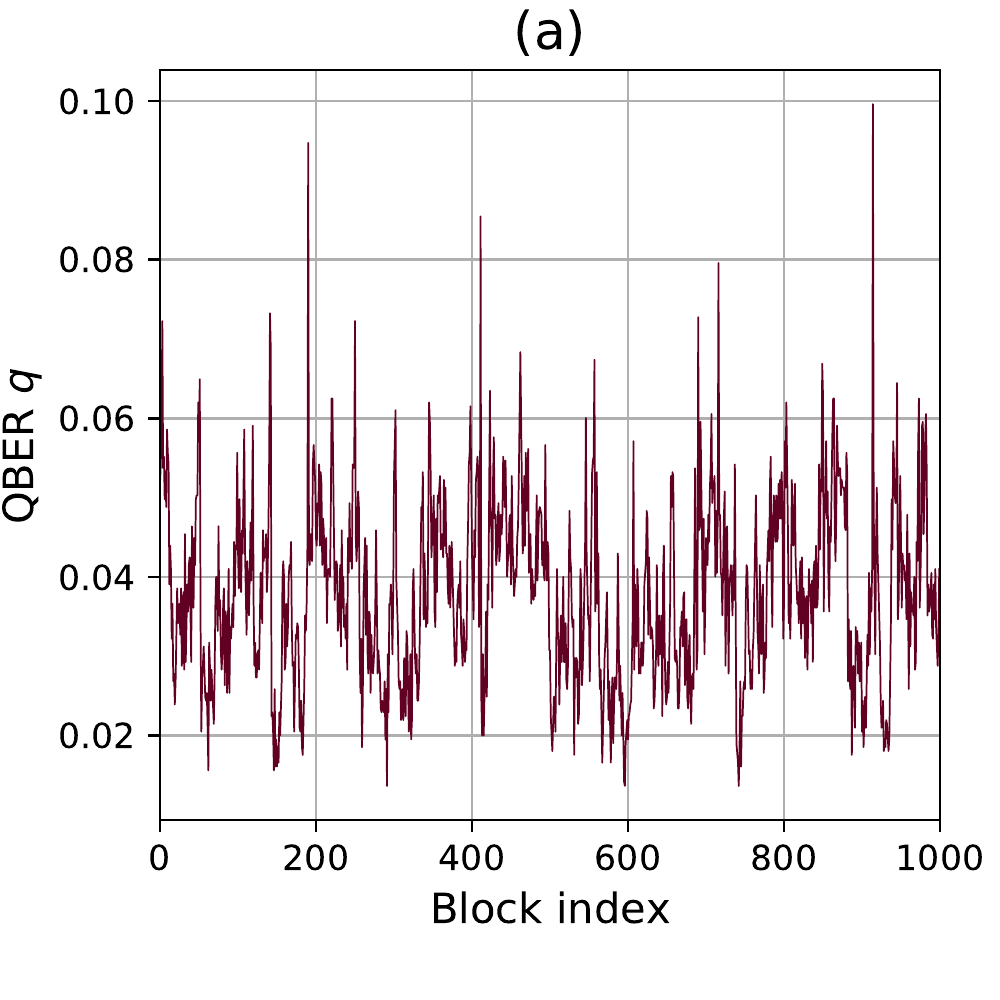}
	\includegraphics[height=0.25\linewidth]{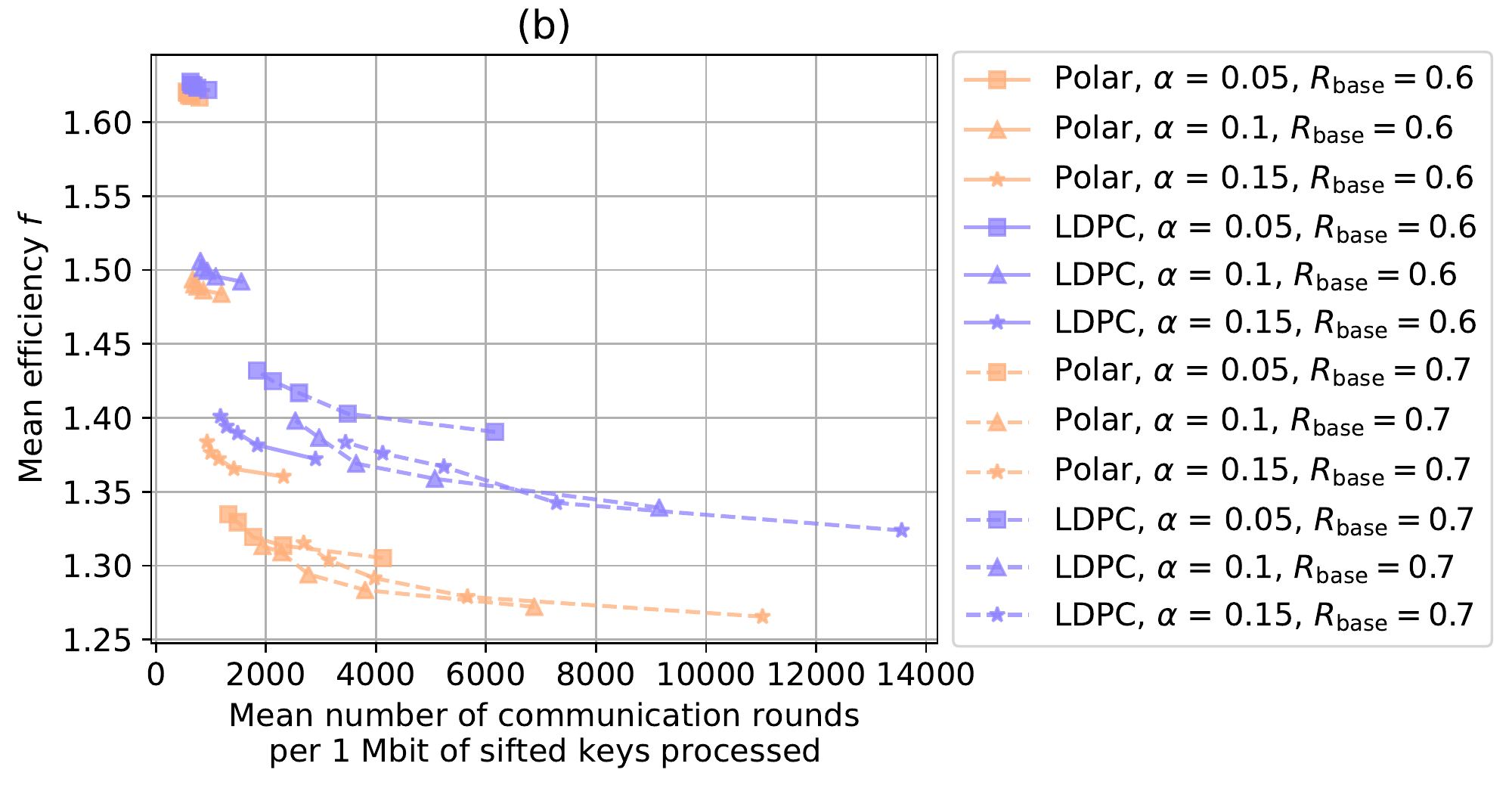}
	\includegraphics[height=0.25\linewidth]{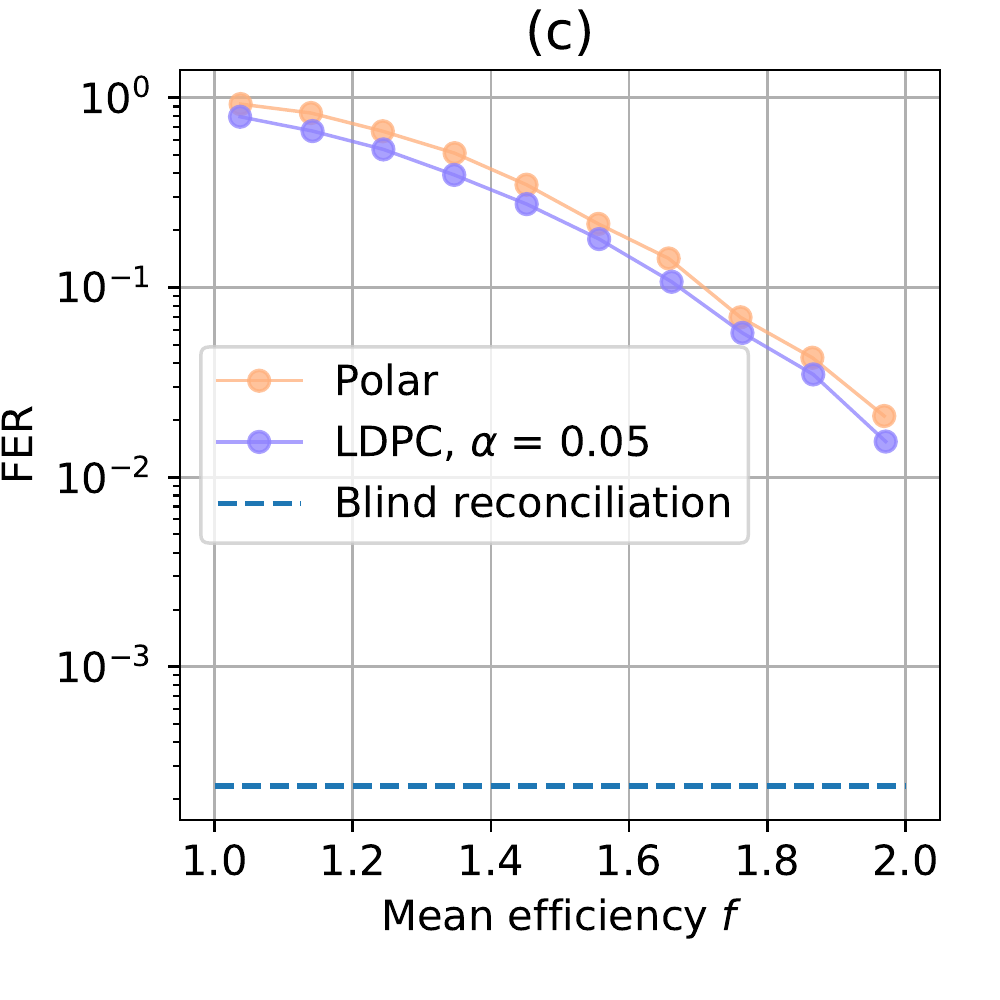}
	\vskip -2mm
	\caption{
		In (a) the behaviour of the QBER level $q$ for the first $10^{3}$ 2048-bit blocks of sifted keys obtained in a real experiment is shown.
		In (b) the tradeoff between a number of communication rounds and efficiency for LDPC and polar codes-based blind reconciliation protocols is demonstrated.
		In (c) the FER as the function of mean efficiency for non-interactive protocols based on LDPC and polar codes 
		{together with a rough estimation of the FER for blind schemes} is depicted.}
	\label{fig:test-2}
\end{figure*}

We use two LDPC codes with rates $\Rbase = 0.6, 0.7$ and consider three different values of $\alpha=0.05, 0.1, 0.15$.
The LDPC codes are constructed with the use of an improved edge-growth algorithm~\cite{Martinez20102} with degree distribution polynomials given by Ref.~\cite{Elkouss2009}.
The frame length is $N=2048$.
To construct polar codes for each pair $(\Rbase, \alpha)$ we choose value of $K$ such that $(K-c)/N=\Rbase/(1-\alpha)$, where the length of a CRC code $c$ is 24 bits.
This choice makes the initial code rates to be the same: $R_{\max}^{\rm LDPC} = R_{\max}^{\rm polar}=:R_{\max}$ [see~Eq.~\eqref{eq:R-range-LDPC} and Eq.~\eqref{eq:R-range-polar}].

{Number of alternative decoding paths $L$ in the Tal-Vardy list decoder was set to 64.}
During the synthetic data test, we generate pairs of sifted keys with fixed QBER $q$.
The performance of two information reconciliation protocols is shown in Fig.~\ref{fig:test-1}.
The parameter $\delta$ for the LDPC codes-based protocol is set to 30, and for the polar codes-based version $\delta$ is set to $\lfloor 30/(1-\alpha)\rceil$ in order to provide the same step with respect to the efficiency [see~\eqref{eq:LDPC-eff} 
and \eqref{eq:polar-eff}].
For the LDPC code-based blind reconciliation protocol Step 5 is replaced with Step 5* and the region, where all punctured bits are spent (actually, where the difference between Step 5 and Step 5* takes place) is highlighted.
We also show values of $f_{R_{\max}}(q) = (1-R_{\max})/h_{\rm bin}(q)$, which correspond to the (theoretical) efficiency of a single-round protocol.

For each pair of $(\Rbase, \alpha)$ we can observe three different types of protocol performance depending on the QBER level.
Let us consider $\Rbase=0.7$, $\alpha=0.1$ as an example.
If QBER is low ($q\leq 0.02$), both LDPC and polar codes cope with reconciliation without additional communication rounds.
Thus their efficiency coincide with $f_{R_{\max}}$.
At moderate QBER values ($0.02<q\leq 0.04$) additional communication rounds start to be required, and both protocols show almost the same performance (one can see that sometimes polar codes show slightly worse performance than LDPC).
If QBER is high ($q>0.4$) and LDPC codes are used, either reconciliation fails (if Step 5 of the original protocol is used), or $f$ and $n$ degrade (if Step 5 is replaced with Step 5*).
Instead, thanks to smooth CDFs shown in Fig.~\ref{fig:CDF}, polar codes demonstrate monotone behaviour approaching some asymptotic value of $f$.
Therefore we can conclude that the use of polar codes is beneficial in the case of high QBER fluctuations, where the third type of behaviour occasionally takes place.

In the second test, we further support this statement and consider a sifted key generated by the industrial QKD devices.
These devices operate in urban conditions, which results in high QBER fluctuations~\cite{Duplinskiy2018}. 
In the Fig.~\ref{fig:test-2}(a) we show the QBER calculated for blocks of 2048 bits.

As in the previous test, we consider LDPC and polar code-based blind reconciliation protocols with Step 5*. 
We use the same set of parameters except for $\delta$ which is varied. 
For the LDPC codes, we take $\delta=(10, 20, 30, 40, 50)$, and for the polar codes we multiply these values by $(1-\alpha)^{-1}$.
Variation of $\delta$ allows one to observe the trade-off between the efficiency and the resulting number of communication rounds.
The results are presented in Fig.~\ref{fig:test-2}(b).
The leftmost (rightmost) point for each curve corresponds to the largest (smallest) step size.
Here we use the mean number of communication rounds per 1 Mbit of sifted key processed as the basic metric.
It emphasizes the fact that with polar codes one processes sifted keys blocks of fixed length.
The plot shows that \textit{for every combination of} $\alpha$ and $R_{\min}$ the polar codes-based protocol outperforms the LDPC codes-based one.

We also provide figures for performance of non-interactive protocols based on LDPC and polar codes.
We take mean value of the QBER $q_{\rm mean}=0.036$ and choose single LDPC and polar code with respect to every target efficiency $f_{\rm target}=1, 1.1, \ldots, 1.9$.
In the case of LDPC codes we use pool of codes with rates $R=0.55, 0.6, \ldots ,0.75$, set $\alpha=0.05$ and then choose the number of punctured bits $p$ and shortened bits
 $s=\alpha N-p$ to have $f_{\rm target} \approx (1-R-p/N)/[(1-\alpha)h_{\rm bin}(q_{\rm mean})]$ (the same frame length $N=2048$ is used).
Since additional communication rounds are not employed, we use fraction of successful reconciliation stages as the main figure of merit.
We show the resulting frame error rate (FER) as a function of resulting efficiency calculated for reconciled blocks in Fig.~\ref{fig:test-2}(c).
We see that for reasonable FERs the resulting (in)efficiency is much higher than the one for the interactive protocols.
To compare, we note that for all tested parameters of interactive protocols we obtain at most one frame error for a number of processed blocks varied from 4250 to 5000 depending on protocol parameters.
So the estimated FER of interactive protocols is of the order of $10^{-4}$.
Also, we see that polar codes demonstrate slightly worse performance than LDPC ones in the non-interactive setting.

\section{Conclusion and outlook}

We have proposed a new interactive protocol for the information reconciliation stage in QKD based on polar codes.
We have shown that even though polar codes show the same or even slightly worth performance compared with LDPC codes in non-interactive protocols, the situation drastically changes in interactive schemes and in the case of high fluctuations of the QBER.
In contrast to the LDPC code-based analogue, the proposed protocol shows consistent gain in successful decoding probability during the process of rate-adaption.
In addition, it processes sifted key blocks of a full-frame length.
These features allow polar codes to outperform LDPC codes in the case of highly unstable QBER level.
Thus, they appear to be a promising alternative to LDPC codes for QKD setups operating in realistic conditions.
We also note that polar codes are of interest for CV-QKD systems, where one-way reverse information reconciliation is a must.
An additional benefit of the suggested protocol is that it uses computational resources of only one communicating side, which opens the perspective for the next generation of QKD networks with star-like topology.
Moreover, additional optimizations and fine-tuning of parameters (e.g. varying of step size in a way as it was implemented for LDPC codes version~\cite{Liu2020}), can be also considered.

\section*{Acknowledgements}

We thank J. Mart{\'{\i}}nez{-}Mateo for useful comments. 
The work is supported by RFBR (Grant No. 18-37-00096, development of the polar codes; Grant No. 18-37-20033, research on the optimization of LDPC codes).

\end{document}